\documentclass[a4paper]{jpconf}
\usepackage{graphicx}
\def\MeV{\nobreak\,\mbox{MeV}}
\def\GeV{\nobreak\,\mbox{GeV}}
\def\beq{\begin{equation}}
\def\enq{\end{equation}}
\def\bea{\begin{eqnarray}}
\def\eea{\end{eqnarray}}

\begin{document}
\title{Exotic Charmonium and Bottomonium-like Resonances}

\author{Fernando S. Navarra$^1$, Marina Nielsen$^1$ and Jean-Marc 
Richard$^2$}

\address{$^1$Instituto de F\'{\i}sica, Universidade de S\~{a}o Paulo,\\
Caixa Postal 66318, 05389-970 S\~{a}o Paulo, SP, Brazil}

\address{$^2$Universit\'e de Lyon and Institut de Physique
Nucl\'eaire de Lyon, IN2P3-CNRS-UCBL \\
4, rue Enrico Fermi, F-69622 Villeurbanne, France}

\ead{navarra@if.usp.br, mnielsen@if.usp.br, j-m.richard@ipnl.in2p3.fr}

\begin{abstract}
Many new states in the charmonium and bottomonium mass region were 
recently discovered  by the BaBar, Belle and CDF Collaborations. We use 
the QCD Sum Rule approach to study the possible structure of some of  
these states. In particular we identify the recently observed  
bottomonium-like resonance $Z_b^+(10610)$ with the first  excitation of 
the tetraquark $X_b(1^{++})$, the analogue of the $X(3872)$ state in the 
charm sector.
\end{abstract}

\section{Introduction}
Most of the new charmonium states discovered in recent years at the 
$B$ factories and at the Tevatron, called $X,~Y,~Z$ particles, do not 
seem to have a simple $c\bar{c}$ structure.  Their
production mechanism, masses, decay widths, spin-parity assignments and decay 
modes have been discussed in some reviews 
\cite{Brambilla:2010cs,Nielsen:2009uh,Olsen:2009gi}.
Although the masses of these
states are above the corresponding thresholds of  decays into a pair of 
open charm mesons, they decay into $J/\psi$ or $\psi^\prime$ plus pions, 
which is unusual for $c\bar{c}$ states. Besides, their masses and decay 
modes are not in agreement with 
the predictions of potential models, which, in general, describe very well
$c\bar{c}$ states. For these reasons, they are considered 
as candidates for exotic states such as  hybrid, molecular or tetraquark 
states, with a more complex structure than the
simple quark-antiquark states. 

In the bottomonium mass region, the first particle that
does not seem to have a simple $b\bar{b}$ structure was the $Y_b(10890)$,
observed by the Belle Collaboration \cite{Chen:2008xia}. The Belle 
Collaboration also reported the observation of two
charged narrow structures in the $\pi^\pm\Upsilon(nS)~(n=1,2,3)$ and
$\pi^\pm h_b(mP)~(m=1,2)$ mass spectra of the $\Upsilon(5S)\to\Upsilon
(nS)\pi^\pm$ and $\Upsilon(5S)\to h_b(mP)\pi^\pm$ decay processes
\cite{bellezb}. These narrow structures were called $Z_b(10610)$ and
$Z_b(10650)$. Analysis of angular distribution favors the quantum numbers
$I^G(J^P)=1^+(1^+)$ for both states. Since the $Z_b$ are charged
states, they are ideal candidates for exotic four-quark states, like the
 $Z^+(4430)$, also observed by  the  Belle Collaboration in $B^+ \rightarrow
K \psi' \pi^+$ through its decay into $\psi^\prime\pi^+$
\cite{Belle:Z4430}. 

\section{Molecular States}

As pointed out by the Belle Collaboration, the proximity 
of the $B\bar{B}^*$ and $B^*\bar{B}^*$ thresholds and the $Z_b(10610)$ and
$Z_b(10650)$ masses suggests that these states could be interpreted as weakly 
bound $B\bar{B}^*$ and $B^*\bar{B}^*$ states
\cite{Bondar:2011ev,Yang:2011rp,nieves,zhu1,meissner}. They have also
been interpreted as cusps at
the $B\bar{B}^*$ and $B^*\bar{B}^*$ thresholds \cite{Bugg:2011jr,russian}
and as tetraquark states \cite{Guo:2011gu}.

The molecular picture was also used to understand the well established
$X(3872)$, which was considered to be a $D\bar D{}^*$ molecule. Of course,
the actual physical state is probably rather complex, with a short-range
component of $(c\bar c)$ nature and a long-range  component with a charmed
meson and an anticharmed meson almost bound by attractive Yukawa forces.
The idea of molecules with hidden charm has been
proposed  long ago by Okun and Voloshin  \cite{Voloshin:1976ap}. A molecular
interpretation was proposed for some high-lying $1^{--}$ charmonium
resonances \cite{De Rujula:1976qd}, due to puzzling branching ratios into 
$D\bar{D}$, $D\bar{D}^*$ and  $D^*\bar{D}^*$, which turned out to be due to the
node structure of these states as radial excitations of the $J/\psi$ 
\cite{Le Yaouanc:1977ux}.
The possibility of hidden-charm meson molecules has been revisited
in the 90s by T\"ornqvist \cite{tornqvist}, Ericson and Karl \cite{ericson}
and Manohar and Wise \cite{manohar}, and further developed by several
authors after the discovery of the $X(3872)$ \cite{Nielsen:2009uh}.
The main idea is that the Yukawa interaction, that successfully binds
nuclei, is not restricted to the nucleon-nucleon interaction. The exchange
of light mesons also generates a potential between flavored mesons, which is
sometimes attractive. Although usually weaker than the proton-neutron
interaction that binds the deuteron, it is probed by much heavier particles,
and thus can lead to bound states with a binding energy of a few MeV, or
even a few tens of MeV, as shown by T\"ornqvist \cite{tornqvist}.
For the molecular $B\bar{B}$ states, the study shows that the energy
of isoscalars $B\bar{B}^*$ with $J^{PC} =0^{-+},~1^{++}$,
$B^*\bar{B}^*$ with $J^{PC}=0^{++},~0^{-+},~1^{+-},~2^{++}$ are
about 50MeV below the corresponding $B\bar{B}^*$ and $B^*\bar{B}^*$
thresholds. No bound state, however, appears for isovectors. Isovector
states receive less attraction than the
states with $I=0$, mainly because of the isospin dependence of the
one-pion-exchange potential. The operator includes a factor $\tau_1.\tau_2$
whose expectation value is $-3$ for $I=0$ and $+1$ for $I=1$. Hence, if the
Belle  $Z_b$ states  are identified as the isovector $B\bar B{}^*$ and
$B^*\bar B{}^*$ molecules, one should also have their isoscalar conterpart
somewhat below (typically a few tens of MeV).

In Ref.~\cite{nieves}, the authors also did not found any isovector
$B\bar{B}^*$ bound state. The only calculation that found loosely bound
states in S-wave $B\bar{B}^*$ and $B^*\bar{B}^*$ for isovectors was
presented by Liu {\it et al.} \cite{zhu1}, where the authors took, besides
the pion, also scalar and vector mesons exchange into account, in the
framework of the meson exchange model.

\section{Tetraquark States}
In the context of the  tetraquark picture,
using the chromomagnetic interaction,
the authors of Ref.~\cite{Guo:2011gu} studied the masses of the
S-wave $[bq][\bar{b}\bar{q}]$ tetraquark states with $J^P=1^+$. They
found six states and  two of them are consistent with the $Z_b(10610)$
and $Z_b(10650)$. However, it is important
to mention that these two states are not the low-lying states in this
channel. The lowest $[bq][\bar{b}\bar{q}]$ tetraquark state
with $J^P=1^+$ appears at 10167.9 MeV. This result is consistent with
the findings of Ref.~\cite{cui} where, using the color-magnetic
interaction with the flavor symmetry breaking corrections, the
$b\bar b q\bar q$ tetraquark states were predicted to be around $10.2
\sim 10.3$ GeV. These results are also consistent with the values
extracted from the QCD sum rule approach \cite{nos,chenwei} for
the $[bq][\bar{b}\bar{q}]$ tetraquark state with $J^{PC}=1^{++}$.

The first QCD sum rule (QCDSR) calculation for the tetraquark
$[b q] [\bar{b}\bar{q}]$ state with $J^{PC}=1^{++}$, which we call $X_b$,
was performed in Ref.~\cite{nos}. At the sum rule
stability point and using the perturbative $\overline{MS}$-mass
$m_b(m_b)=4.24~\GeV$, the authors obtained $M_{X_b}=   (10250\pm200)~\MeV$,
for  $\sqrt{s_0}=(10500\pm300)~\MeV $, where $s_0$ is the continuum
threshold. The authors of  Ref.~\cite{chenwei} have used different
$J^{PC}=1^{++}$ and $J^{PC}=1^{+-}$ tetraquark $[bq][\bar{b}\bar{q}]$
currents. They have obtained $M_{X_b}=   (10220\pm100)~\MeV$,
for  $\sqrt{s_0}=(10800\pm100)~\MeV $ which is in complete agreement
with the result of  Ref.~\cite{nos}. The authors of Ref.~\cite{Zhang:2011jj} 
tried to reproduce the mass of $Z_b(10610)$
using a $B\bar{B}^*$ molecular current in the QCDSR calculation.
They obtained  $M_{BB^*}=   (10560\pm180)~\MeV$, which they say is in
agreement with the $Z_b(10610)$ mass. However, to obtain such a big mass
they were forced to use a continuum threshold of $\sqrt{s_0}=(11400\pm200)
~\MeV $, which is much bigger than the values used in
Refs.~\cite{nos,chenwei}.
As it was shown in Ref.~\cite{drx}, different currents with the same
quantum numbers lead to approximately the same mass in the QCDSR approach,
if the same parameters are used. Therefore, from  QCDSR studies one may
say that the low-lying $X_b$ state
has a mass around 10100 - 10200 MeV, which is in agreement with the
results from chromomagnetic model calculations \cite{Guo:2011gu,cui}.

It is very interesting to notice that the mass difference between
the predicted $X_b$ and $\chi_{b1}(9892)$:
\beq
M_{X_b}-M_{\chi_{b1}}\sim 310~\MeV,
\label{difb}
\enq
is of the same order of magnitude of the mass difference between the
$X(3872)$ and $\chi_{c1}(3510)$:
\beq
M_{X}-M_{\chi_{c1}}\sim 360~\MeV.
\label{difc}
\enq
This kind of similarity between the $c$ and $b$ sector is very interesting.
One can see that $M_{\Psi(2S)}-M_{\Psi(1S)}=590~\MeV\sim
M_{\Upsilon(2S)}-M_{\Upsilon(1S)}=560~\MeV$. Therefore, the results in
Eqs.~(\ref{difb}) and (\ref{difc}), could be used as an evidence that
a 4-quark state, similar to $X(3872)$,  should  exist in the $b$ sector!
This state should be searched for by the experimental groups.

We suggest that the recently observed $Z_b$ states are not
ground states of the $1^+$ bottomonium 4-quark states, but
excitations of a ground state with a mass around $10.2$ GeV. A similar
suggestion was  made by Maiani, Polosa and Riquer \cite{maiani},
to explain the $Z^+(4430)$  as an excitation of a charged $X$ state.
Although the Babar Collaboration found no conclusive evidence of the
existence of $Z^+(4430)$ \cite{babarz}, Belle has confirmed its observation.
Using the same data sample of  Ref.~\cite{Belle:Z4430}, Belle also
performed a full Dalitz plot analysis \cite{bellezn} and  has confirmed
the observation of the $Z^+(4430)$ signal with a 6.4$ \, \sigma$ peak
significance. Therefore, if one believes in the $Z_b$ states
observed by Belle, one should also believe in the existence of the
$Z^+(4430)$.

\section{$Z_b$ is not $X_b$. Is that a question?} 

In \cite{maiani2} it was conjectured that the $X(3872)$
must have a charged partner $X^+$  with $J^{PC}= 1^{+-}$ with a similar
mass. In \cite{maiani} it was pointed out that since the
mass difference
\beq
M_{Z^+ (4430)} -  M_{X^+(3872)}\sim 560 \, \MeV
\label{difz}
\enq
is close to the mass difference $M_{\psi(2S)} - M_{\psi(1S)}$ given
above, the  $Z^+(4430)$  may well be the
first radial excitation of the $X^+$. In a straightforward extension
of this reasoning to the bottom case,
we may conjecture that the $Z_b (10610)$ is also a radial excitation
of an yet unmeasured $X_b^+$ ($J^{P}= 1^{+}$)  state, such that the
mass difference
\beq
M_{Z_b(10610)}- M_{X_b^+(10100)} \sim 510 \, \MeV
\label{difzb}
\enq
very  close to the  mass difference $M_{\Upsilon(2S)} - M_{\Upsilon(1S)}
 = 560 \, \MeV$. For the sake of clarity the above mentioned numbers are
displayed in Fig.~1, where we compare the charm and bottom spectra in the
mass region of interest. On the left (right) we show the charm (bottom)
states with their mass differences in MeV. The comparison between the two
left lines with the two lines on the right emphasizes the similiraty between
the conjecture made in this note, namely the existence of $X^+_b$ as a ground
state of the measured $Z_b^+ (1610)$, and the conjecture made in Ref.
\cite{maiani} on the existence of  the charged partner $X^+$ of the
$X(3872)$ as the ground states of the  $Z^+(4430)$.

\begin{figure}[h]
\centering\includegraphics[width=35pc]{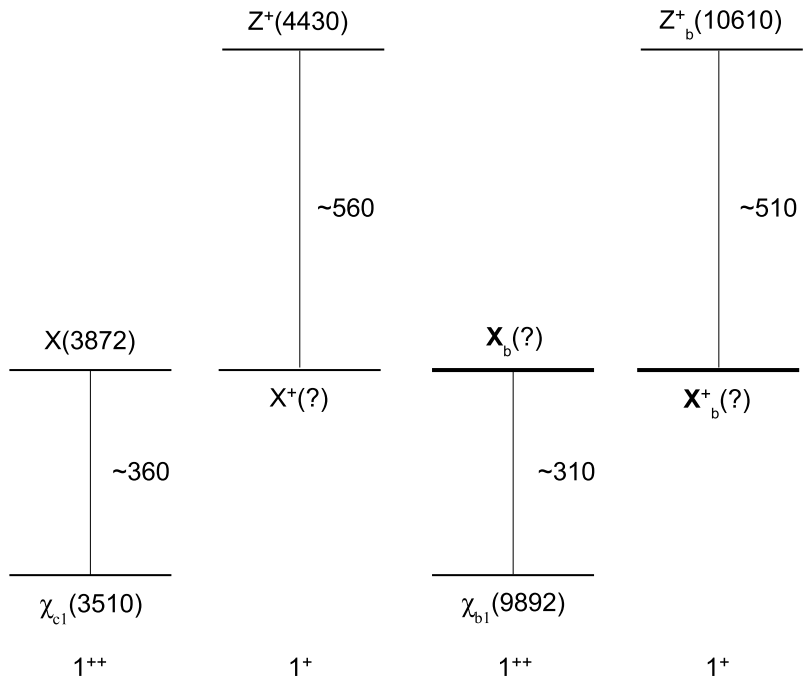}
\caption{\label{fig1}Charm and bottom energy levels in the mass region of 
interest. Masses are in MeV. On the
two left columns we show the conjecture presented in \cite{maiani}. 
The $X^+ (?)$ 
is the proposed charged partner of the $X(3872)$. 
On the two right columns we show our equivalent conjecture for the bottom
sector, with $X_b(?)$ and $X_b^+(?)$ being the proposed states.}
\end{figure}

The conjecture presented above should encourage searches both in the charm
and bottom sectors in the mass regions  around  $3870$ MeV and $10200$ MeV 
respectively.
The reason for these searches would be  not only to find the unobserved states
shown in Fig.~1, $X^+(?)$, $X_b(?)$ and $X^+_b(?)$.
 In fact there are even more
striking states to be detected in these mass regions. In the molecular
approach, there is also attraction in the flavor exotic $DD$ or $D^*D^*$ and
$BB^*$ or $B^*B^*$ channels. In the vector-vector case, the Fermi--Yang rule
of $G$-parity \cite{foot} holds and
the pion-exchange potential flips sign (for a given isospin $I$). Hence
$B^*\bar B^*$ channels with repulsive interaction are transformed into
$B^*B^*$  with attraction. The pseudoscalar-vector case is more subtle.
Pion-exchange induces a transition from $|1\rangle=B\bar B^*$ to
$|2\rangle=B^*\bar B$, and by linear combination, the potential is
attractive in one of the channels, say $|1\rangle\pm |2\rangle$.  In the
flavor-exotic sector, one deals with $|1'\rangle=B B^*$ and $|2'\rangle=B^*
B$. The pion-exchange potential still flips signs,
but this simply means that the very same attraction is observed now in
the combination $|1'\rangle\mp |2'\rangle$, if it exists in the partial wave one looks at.

For tetraquark systems, constituent model calculations have always favored
$(QQ\bar q\bar q)$ configurations, where the pair of heavy quarks benefits
from their attraction. In the threshold $(Q\bar q)+(Q\bar q)$, instead, the
heavy quarks interact only with light quarks, and no heavy reduced mass
enters. For a review, see, e.g., \cite{Vijande:2009xx}. This peculiarity 
of $(QQ\bar q\bar
q)$ channels is confirmed in some studies based on lattice QCD 
\cite{Michael:1999nq} and QCD
sum rules \cite{Navarra:2007yw}. 
Experimentally the sector with charm $+2$ or beauty $-2$ is
almost virgin. Theoretically, the question is whether charm is heavy enough
to make $(cc\bar q\bar q)$ stable against spontaneous dissociation or $b$
quarks are necessary. Two $b$ or not two $b$, that is the question!

\section{Conclusions}

We have discussed the masses of some $X$, $Y$ and $Z$ states, recently 
observed by BaBar and Belle Collaborations. In some cases a tetraquark 
configuration was favored, as the $Y_b(10890)$ \cite{Albuquerque}, 
and in some other cases a molecular configuration was favored. 
In the case of $Z_b^+(10610)$ we identify it as
the first  excitation of the tetraquark $X_b(1^{++})$, the analogue of
the $X(3872)$ state in the charm sector.

\section*{Acknowledgments}
This work has been partly supported
by the CNRS-FAPESP program,  by  CNPq-Brazil and by the CNRS-IN2P3
within the project Non-perturbative QCD and Hadron Physics. We thank Frank Close, Christopher Thomas and Nils T\"ornqvist for a (still ongoing) lively discussion about the molecular dynamics.


\begin{thebibliography}{9}
\bibitem{foot} It was much used in the time of baryonium to relate
the nucleon--antinucleon interaction to the nucleon-nucleon one.
\end{thebibliography}
\end{document}